\documentclass[prl,%
reprint,
 amsmath,amssymb,
 aps,
]{revtex4-1}

\usepackage{graphicx}
\usepackage{dcolumn}
\usepackage{bm}

\begin{document}

\preprint{PRL}

\title{Solid-liquid composite structures: elastic beams with embedded liquid-filled parallel-channel networks}

\author{Y. Matia}
\affiliation{Faculty of Mechanical Engineering, Technion - Israel Institute of Technology,
Technion City, Haifa, Israel 32000}
\author{A.D. Gat}%
\email{amirgat@tx.technion.ac.il}
\affiliation{Faculty of Mechanical Engineering, Technion - Israel Institute of Technology,
Technion City, Haifa, Israel 32000}
\date{\today}

\begin{abstract}
Deformation due to embedded fluidic networks is currently studied in the context of soft-actuators and soft-robotics. Expanding on this concept, beams can be designed so that the pressure in the channel-network is created directly from external forces acting on the beam, and thus can be viewed as passive solid-liquid composite structure. We obtain a continuous function relating the network geometry to the deformation. This enables design of networks creating arbitrary steady and time varying deformation-fields as well as to eliminate deformation created by external forces.
\end{abstract}

\maketitle
Fluid enclosed within an elastic solid may apply pressure and shear stress on the fluid-solid interface and thus create a stress-field and a deformation-field within the solid \cite{Heil.1995,Ku.1997,Paidoussis.1998,Canic.2003}. The interaction between the pressure-field of an internal fluid-filled channel network on the deformation-field of the supporting elastic structure is currently researched within the context of soft robotics and soft actuators \cite{Shepherd.2011,Martinez.2013,steltz2009jsel,Stokes.2013,Rus.2014,majidi2014soft,Elbaz2014}. In this work we expand on the concept of pressurized soft actuators and suggest utilizing pressurized parallel-channel networks to significantly increase the effective rigidity of elastic structures by canceling deformation-fields created by steady or time varying external forces. The pressure within the channel network can be applied directly by the external forces (e.g. by a pin in contact with the liquid) and the structure thus can be viewed as a passive solid-liquid composite structure. 

\begin{figure}
\includegraphics[width=.45\textwidth]{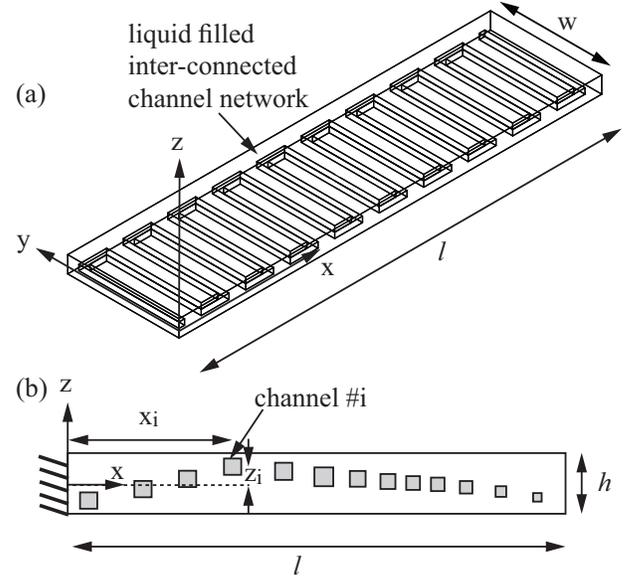}
\caption{A three-dimensional illustration (a) and a cross section illustration (b) of an elastic beam with an embedded interconnected parallel channel network.}
\label{figure_1}
\end{figure}

We focus on a rectangular beam with height $h$, width $w$ and length $l$ under the requirements $h/w\ll1$ and $w/l\ll1$ (see Fig. \ref{figure_1}a). The Young's modulus, density, Poisson's ratio and mass per unit length of the beam are $E$, $\rho$, $\nu$ and $\mu_s$, respectively.  An interconnected parallel channel network is distributed within the beam perpendicular to the $x-z$ plane (see Fig. \ref{figure_1}b). The liquid pressure is $p$ and is assumed spatially uniform. The difference between the length of a single channel and the width of the beam $w$ is required to be negligible compared with $w$. The total length of channel segments connecting the parallel channels is required to be negligible compared with the total length of channel network. In addition, we focus on channel networks with negligible effect on the second moment of inertia and mass per unit length of the beam. The deflection of the beam in the $z$ direction is denoted by $d$. We assume small deformations so that $d=d_e+d_n$ is a sum of $d_e$, the deformation due to external forces, and $d_n$, the deformation due to the pressurized channel network.

\begin{figure}
\includegraphics[width=.45\textwidth]{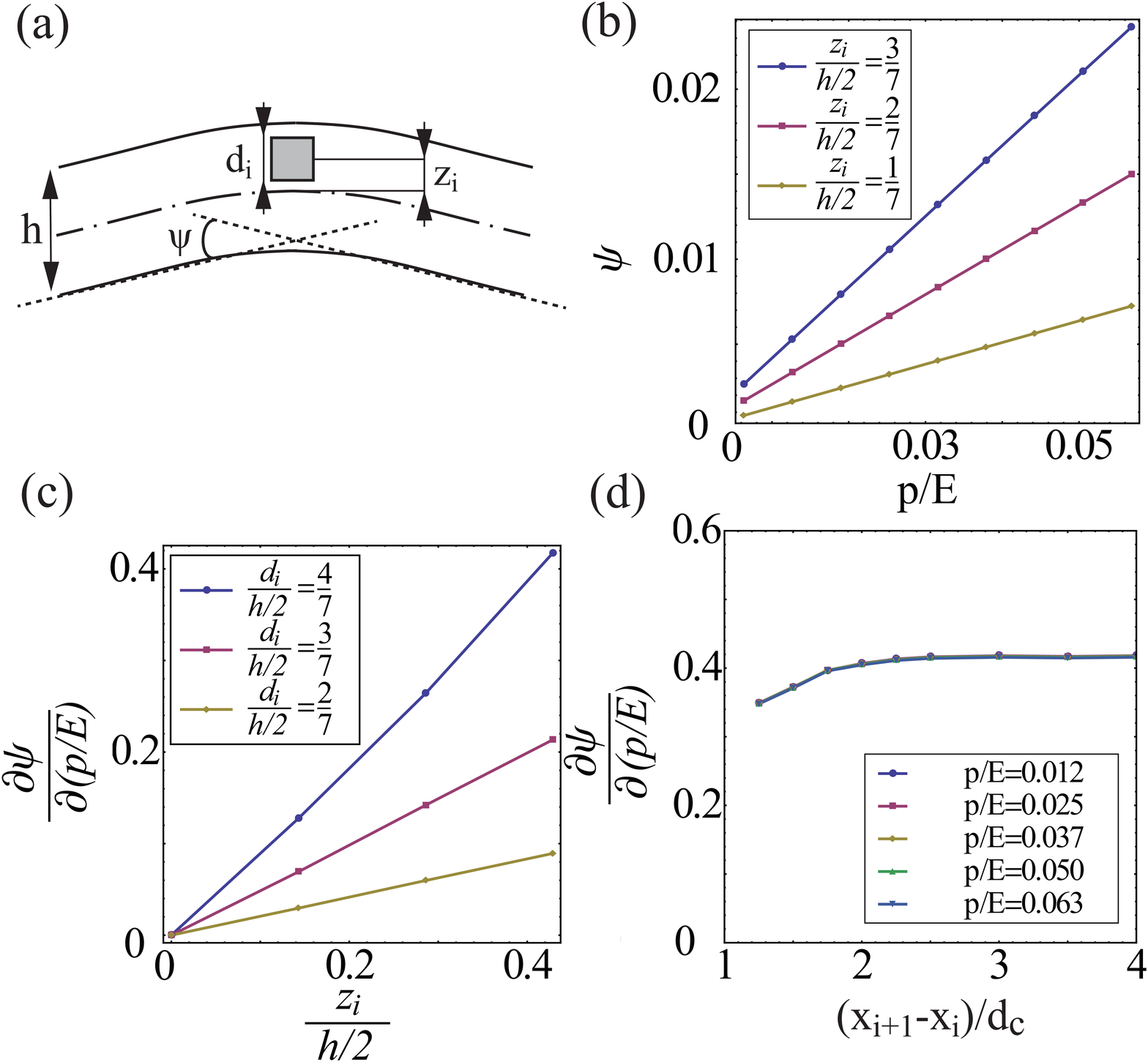}
\caption{(a) The definition of $\psi$ and the geometric parameters of the channel. 
(b) $\psi$ vs. $p/E$ for various values of $z_i/(h/2)$ where ${d_i}/{(h/2)}={4}/{7}$. (c) $\partial \psi /\partial (p/E)$ vs. $z_i/(h/2)$ for various $d_i/(h/2)$. (d)  $\partial \psi /\partial (p/E)$ vs. $(x_{i+1}-x_i)/d_i$, the distance between centers of adjacent channels, for various $p/E$. In (b-c) the channel cross-section is a square with width and height $d_i$.}
\label{figure_2}
\end{figure}

A single pressurized channel will create local stress and strain-fields which will decay far from the channel \cite{dugdale1971elasticity}. A pressurized channel positioned asymmetrically with regard to the mid-plane will create a change of the slope of the beam due to asymmetric strain-field (see Fig. \ref{figure_2}a). For a sufficiently small ratio $h/w\ll1$, the problem is approximately two-dimensional and thus we can define the change in beam slope due to a single channel as $\psi$
\begin{equation}
\frac{\partial d_n(x_i+\Delta x)}{\partial x}-\frac{\partial d_n(x_i-\Delta x)}{\partial x}=\psi\left(\frac{p}{E},\nu,\frac{z_i}{h},\frac{d_i}{h}\right),
\end{equation}
where $x_i$ is the location of the center of the channel and $\Delta x$ is sufficiently large so that the stress-field vanishes. The value of $\psi$, the change in beam slope due to a single channel, can be obtained numerically or experimentally for a given material, pressure and channel configuration. In Fig. \ref{figure_2} we present values of $\psi$ obtained by numerical computations for a channel with a square cross-section. Fig. \ref{figure_2}a illustrates the definition of $\psi$ and the geometric parameters of the channel, including $z_i$, the distance of the channel center from the midplane, and $d_i$, the width and height of the square cross-section. Fig. \ref{figure_2}b presents $\psi$ vs. $p/E$ for various values of $z_i/(h/2)$ where ${d_i}/{(h/2)}={4}/{7}$. Fig. \ref{figure_2}c presents $\partial \psi /\partial (p/E)$ vs. $z_i/(h/2)$ for various $d_i/(h/2)$. Fig. \ref{figure_2}b,c show that $\psi$ increases monotonically with $z_i/(h/2)$ and $d_i/(h/2)$. Fig. \ref{figure_2}d presents $\partial \psi /\partial (p/E)$ vs. $(x_{i+1}-x_i)/d_i$ for various $p/E$, examining the effect of interaction between adjacent channels on $\partial \psi /\partial (p/E)$. The influence of adjacent channels is shown to be small, even for distances of $(x_{i+1}-x_i)/d_c\approx 1.2$. From Fig. \ref{figure_2}a-c, the value of $\psi$ is approximately linear with $p/E$ and thus,
\begin{equation}
\psi \approx \frac{p}{E}\frac{\partial\psi}{\partial (p/E)}\left(\frac{p}{E}=0,\nu,\frac{z_i}{h},\frac{d_i}{h}\right).
\label{psi_eq}
\end{equation}

We define the channel density $\phi$ of a parallel channel network (see Fig. \ref{figure_1}) as the number of channels per unit length. For characteristic length scale $l$ much greater than the characteristic distance between the channels ($l\gg 1/\phi$) we can approximate the change in slope to a continuous function
\begin{equation}
\frac{\partial^2 d_n}{\partial x^2 }=\frac{1}{dx}\left(\frac{\partial d_n(x+dx)}{\partial x}-\frac{\partial d_n(x)}{\partial x}\right)=\frac{1}{dx}\left(k \psi \right),
\label{phi_eq_0}
\end{equation}
where $k$ is the number of channels in the interval $dx$. Defining the local density of the channels as $\phi=k/dx$ and applying (\ref{psi_eq}) yields a relation between the parallel channel configuration and the deformation pattern created by the pressurized network, denoted as $d_n$,
\begin{equation}
\frac{\partial^2 d_n}{\partial x^2 }=-\phi\frac{p}{E}\frac{\partial\psi}{\partial (p/E)}\left(\frac{p}{E}=0,\nu,\frac{z_i}{h},\frac{d_i}{h}\right).
\label{phi_eq}
\end{equation}

Throughout this work we present numerical computations in order to validate our analysis. In all cases we simulate a beam with $h=7\cdot 10^{-3}m$, $w=5\cdot 10^{-2}m$, $l=0.1m$, $E=8 \cdot 10^6Pa$, $\rho=1100Kg/m^3$ and $\nu=0.4$. The channel cross section is square with width $d_i/(h/2)=4/7$. The beam includes a $0.5mm$ area on all sides without a network and the connecting channels have identical properties to the parallel channels. A spatially uniform pressure is applied at the solid-liquid interface. Our computations utilize commercial code COMSOL multiphysics 4.3 with $\approx10^5$ grid elements to calculate the solid deformation.

From (\ref{phi_eq}) we can obtain the required geometry of a channel network to create a predetermined deformation field $d_n$. After calculating $\phi$ from Eq. (\ref{phi_eq}) the location of the center of the channel $x_i$ is determined by 
\begin{equation}
\int_0^{x_i}{|\phi| dx}=i-\frac{1}{2},
\label{xi_eq}
\end{equation}
where $i$ is a natural number. Hereafter, in all cases, we solve $\phi$ for $z_i/(h/2)=4/7$. For cases in which we obtain $\phi<0$ (negative channel density) we replace $z_i/(h/2)=4/7$ with $z_i/(h/2)=-4/7$ and thus change the sign of $\partial \psi/\partial(p/E)$.

\begin{figure}
\includegraphics[width=.45\textwidth]{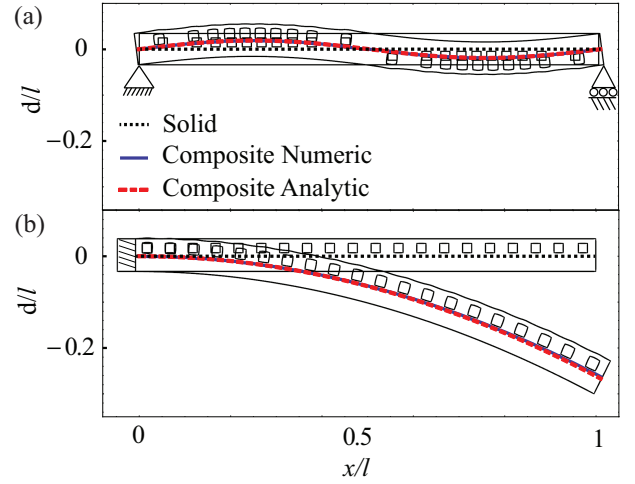}
\caption{The deformation field created by a channel network calculated by Eq. (\ref{phi_eq}) for $d_n/l=0.02\sin(2 \pi x /l)$ (a) and $d_n/l=2+\sqrt{4-(x/l)^2}$ (b). Composite solid liquid deflection is marked by red dashed lined (analytic) and smooth blue lines (numeric). For comparison, deformation-field without the network is marked by black dotted lines.}
\label{figure_3}
\end{figure}

Fig. \ref{figure_3} illustrates the creation of an arbitrary steady deformation field of the beam by designing the channel network according to Eq. (\ref{eli_def}). Fig. \ref{figure_3}a presents sine deformation field $d_n/l=0.05\sin(2 \pi x/l)$ and Fig. \ref{figure_3}b presents a circular deformation defined by $(x/l)^2+(d_n/l+2)^2=4$. Good agreement is observed between the model (red dashed lines) and numerical computations (blue solid lines).

For slender linearly elastic beam the deformation created by steady external forces, denoted as $d_e$, is given by Euler-Bernoulli beam theory as $\partial ^2 d_e/\partial x^2=M/EI$, where $M$ is the bending moment and $I=h^3w/12$ is the second moment of inertia. Assuming small deformations, the total deflection of the beam is $d=d_n+d_e$. Thus, the deflection due to external forces, $d_e$, can eliminated by requiring
\begin{equation}
\frac{\partial ^2 d_n}{\partial x^2}+\frac{\partial ^2 d_e}{\partial x^2}=0\rightarrow  -\frac{p(t)}{E}\phi(x)\frac{\partial \psi(x)}{\partial (p/E)}+\frac{M}{EI}=0.
\label{eli_def}
\end{equation}
Therefore, for any bending moment distribution which can be presented as $M=f_1(t)f_2(x)$ the deflection field can be eliminated by requiring $p(t)=f_1(t)$ and $\phi(x)\partial \psi(x)/\partial (p/E)=f_2(x)$. Since the total deformation $d=d_e+d_n$ is constant, no inertial effects will be created due to the time-varying external forces. 

\begin{figure}
\includegraphics[width=.45\textwidth]{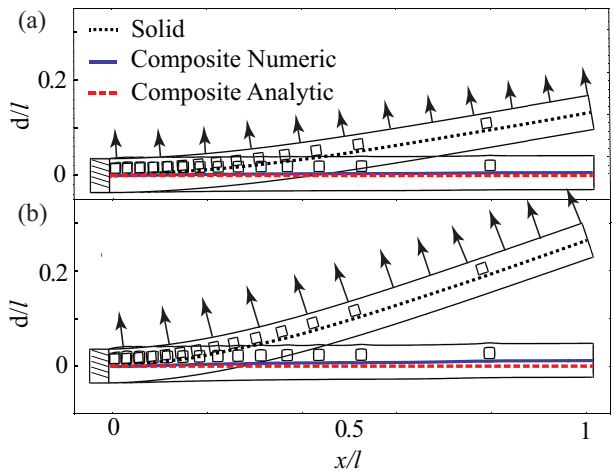}
\caption{The deformation field created by a channel network calculated by Eq. (\ref{eli_def}) in order to cancel external uniform load. Two values of $q/E=2.5\cdot10^{-5}$ (a) and $q/E=5\cdot10^{-5}$ (b). Composite solid liquid deflection is marked by red dashed lined (analytic) and smooth blue lines (numeric). For comparison, deformation-field without the network is marked by black dotted lines.}
\label{figure_4}
\end{figure}

In Fig. \ref{figure_4} we illustrate utilizing an internal fluidic network to enhance the effective rigidity of an elastic beam. The required deformation-field is marked by red dashed lines and the deformation obtained by numerical computations is marked by solid blue lines. For comparison a solid beam without embedded channel network is presented by dotted black lines. For the case of uniform load $q/E=2.5\cdot10^{-5}$ (e.g. load acting on a wing) the required network geometry was calculated according to Eq. (\ref{eli_def}) for $p/E=3.16\cdot 10^{-2}$. Since the deformation is linear both with $p/E$ and with $q/E$, an increase of the external load to $q/E=5\cdot10^{-5}$ would be eliminated by a proportional increase in the network pressure to $p/E=6.32\cdot 10^{-2}$. Thus, the cancellation of deformation by varying external load can be eliminated by a single network configuration.

\begin{figure}
\includegraphics[width=.45\textwidth]{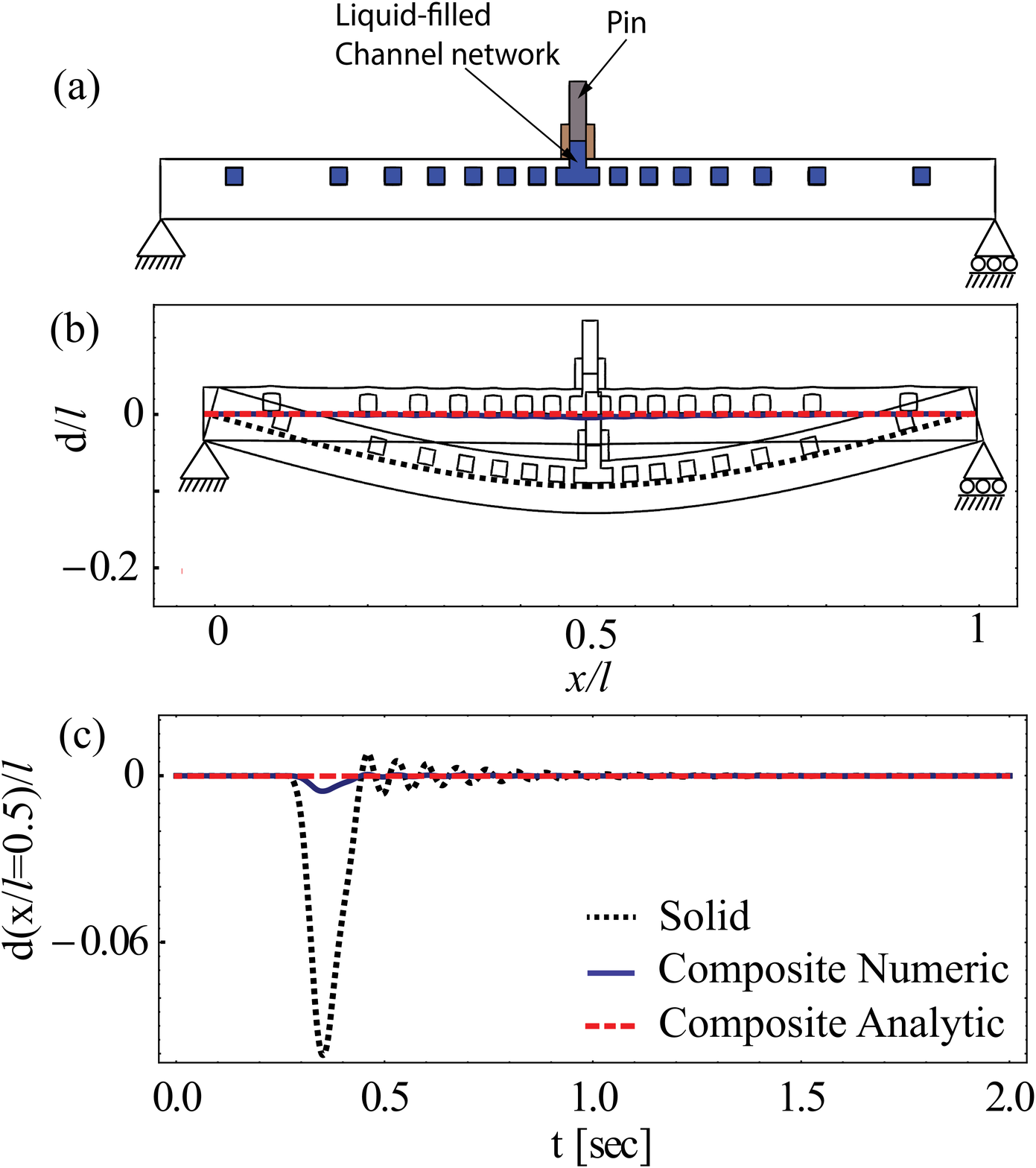}
\caption{(a) A cross section illustration of a solid-liquid composite structure. (b) The deformation field created by the external force $f$ on the pin (where $f=3.15[N]$ and $a=4\cdot 10^{-6}[m^2]$). (c) Response to an impulse $f=\delta(t-t_s)0.63[N]$ where $t_s=0.35[s]$. The channel network calculated by (\ref{phi_eq}) in order to eliminate deformation. The deformation obtained by numerical computations of the solid deformation is marked by solid blue lines.}
\label{figure_5}
\end{figure}

Based on the above we suggest a liquid-solid composite structure in which application of external force directly creates pressure within an internal channel network. Such structures will allow to control the dynamic response of beams to external loads by the addition of the deformation created by the pressurized network configuration to the deformation created by external forces. An illustration of such a structure is presented in Fig. \ref{figure_5}a where a force $f$ may be applied by a pin directly on the liquid, creating a liquid pressure $p=f/a$, where $a$ is the area of the pin. Fig. \ref{figure_5} presents the response of such a structure to steady external force $f=0.2[N]$ (Fig. \ref{figure_5}b) and to a sudden impulse $f=\delta(t-t_s)0.63[N]$, where $\delta$ is Dirac's delta function  (Fig. \ref{figure_5}b). Order of magnitude reduction in deformation is observed for both the steady and time-varying external forces.

So far we focused on creating steady deformation fields. In order to create a pre-defined time varying deformation field the design of the internal channel network will need to include the effect of solid inertia. The deformation field created by the channel-network yields acceleration of the beam and thus the Euler-Bernoulli equation is
\begin{equation}
\frac{\partial ^2 }{\partial x^2}\left(EI \frac{\partial ^2 d_e}{\partial x^2}\right)=-\mu_s\frac{\partial ^2}{\partial t^2}\left(d_e+d_n\right)+qw,
\end{equation}
where $\mu_s$ is beam mass per unit length. Substituting $d=d_e+d_n$ and Eq. (\ref{phi_eq}) yields equation of the total deflection including the effects of the channel network geometry and time varying (spatially uniform) pressure as
\begin{equation}
\frac{\partial ^2 }{\partial x^2}\left[EI \left(\frac{\partial ^2 d}{\partial x^2}+\phi \frac{p}{E} \frac{\partial \psi}{\partial (p/E)}\right)\right]=-\mu_s\frac{\partial ^2 d}{\partial t^2}+qw.
\label{gov_time_Eq}
\end{equation}

Solution of Eq. (\ref{gov_time_Eq}) can be obtain for an oscillating deformation of the form 
$d/l=D (x)\sin(\omega t+\theta)$ under similarly oscillating external load $q=Q(x)\sin(\omega t+\theta)$, where $D(x)$ and $Q(x)$ are known functions defining deformation and external load, respectively, $\omega$ is the angular frequency and $\theta$ is the phase. For the case of a solid-liquid composite (see Fig. \ref{figure_5}a) the internal pressure is proportional to the external force and thus $p=P\sin(\omega t)$, where $P$ is a known constant. Substituting $d/l$, $q$ and $p$ into Eq. (\ref{gov_time_Eq}) yields the required network density,
\begin{eqnarray}
\phi=\left(P \frac{\partial \psi}{\partial (p/E)}EI\right)^{-1}\times\quad\quad\quad\quad\quad\quad\quad\quad\quad\quad
\label{phi_oci}
\\
\bigg[\int_0^x{\int_0^\eta{\left(\mu_s\omega_n^2 l D(\xi)-wQ(\xi)\right)}d\xi}d\eta-l\frac{\partial ^2 D}{\partial x^2}\bigg]. \nonumber
\end{eqnarray}

\begin{figure}
\includegraphics[width=.45\textwidth]{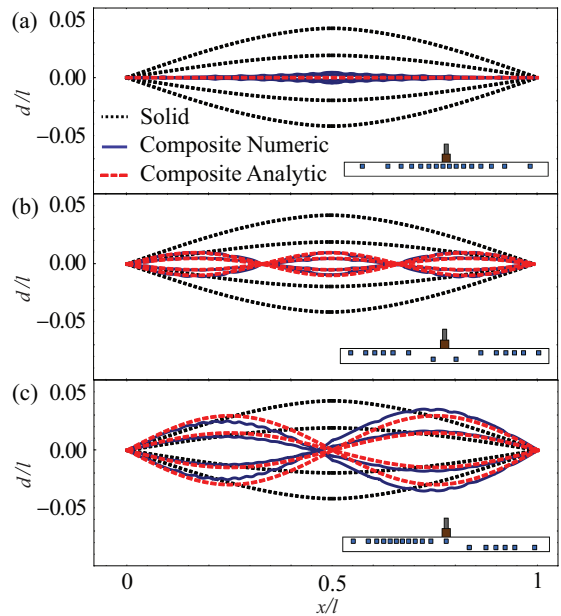}
\caption{Deflection of a solid-liquid composite beam due to external oscillating force acting at $x/l=0.5$. The parallel channel network (illustrated in inserts) is designed by Eq. (\ref{phi_oci}) to create deflection (a) $d/l=0$, (b) $d/l=0.01\sin(3\pi x/l)\sin(\omega t)$ and (c) $d/l=0.03\sin(2\pi x/l)\sin(\omega t)$, where $\omega=62.8[1/s]$. Each time cycle is divided to four equal parts. Composite solid liquid deflection is marked by red dashed lined (analytic) and smooth blue lines (numeric). For comparison, deformation-field without the network is marked by black dotted lines.}
\label{figure_6}
\end{figure}

We illustrate use of (\ref{phi_oci}) for the case presented in Fig. \ref{figure_5}a with $q/E=C_1 \delta(x/l-1/2) \sin(\omega t)/w$ and thus $p/E=C_1 \sin(\omega t) w l/a$, where $a=2.207\cdot 10^{-6}[m^2]$ is the area of the pin and $C_1=2[N]$. The value of $\omega$ is $62.8[1/s]$, where the natural angular frequency of the beam is $\approx88[1/s]$. Fig. \ref{figure_6}a presents the effects of oscillating external force for $\phi$ designed by Eq. (\ref{phi_oci}). In part (a) $d/l=0$, in part (b) $d/l=0.01\sin(3\pi x/l)\sin(\omega t)$ and in part (c) $d/l=0.03\sin(2\pi x/l)\sin(\omega t)$. The required deformation-field is marked by red dashed lines and the deformation obtained by numerical computations is marked by solid blue lines. Good agreement is observed between the theoretic predictions and the numerical computations.

Concluding, embedded fluidic networks can create complex time-varying deformation patterns in elastic beams. By utilizing external forces to directly pressurize the channel-network (as presented in \ref{figure_5}a) composite solid-liquid structures, with unique mechanical properties, can be designed. Such structures can be relevant to soft-actuators, soft-robotics and aerospace structures. Future research may include the effects of liquid viscosity and a non-uniform pressure distribution on the transient response of such structures to external forces.

\bibliography{Bib_File}

\end{document}